\documentclass[prd,aps,floats,preprintnumbers,preprint]{revtex4}
\pdfoutput=1
\usepackage{graphicx}
\usepackage[english]{babel} % grammar package
\usepackage{braket}%Dirac notation
\newcommand{\be}{\begin{equation}}
\newcommand{\ee}{\end{equation}}
\newcommand{\bea}{\begin{eqnarray}}
\newcommand{\eea}{\end{eqnarray}}

\newcommand{\eqn}[1]{(\ref{#1})}

\usepackage{morefloats} %When there are many figs.~
\usepackage[titletoc,title]{appendix}
\usepackage{color}
\begin{document}

\title{Degeneracy in the spectrum and bispectrum among featured inflaton potentials}

\author{Alexander Gallego Cadavid$^{1,2,3,4}$, Antonio Enea Romano$^{1,2,3}$, Misao Sasaki$^{3}$}
\affiliation
{
${}^{1}$ICRANet, Piazza della Repubblica 10, I--65122 Pescara, Italy \\
${}^{2}$Instituto de F\'isica, Universidad de Antioquia, A.A.1226, Medell\'in, Colombia\\
${}^{3}$Center for Graviational Physics, 
Yukawa Institute for Theoretical Physics, Kyoto University, Japan \\
${}^{4}$Exact and Applied Science, Instituto Tecnológico Metropolitano, Street 73 76A-354, Medellín, Colombia
}
\begin{abstract}
We study the degeneracy of the primordial spectrum and bispectrum of the curvature perturbation in single field inflationary models with a class of features in the inflaton potential. The feature we consider is a discontinuous change in the shape of the potential and is controlled by a couple of parameters that describe the strength of
the discontinuity and the change in the potential shape. This feature produces oscillations of the spectrum and bispectrum around the comoving scale $k=k_0$ that exits the horizon when the inflaton passes the discontinuity.
We find that the effects on the spectrum and almost all configurations of the bispectrum including the squeezed limit depend on a single quantity which is a function of the two parameters defining the feature. This leads to a degeneracy, i.e. different features of the inflaton potential can produce the same observational effects. However, we find that the degeneracy in the bispectrum is removed at the equilateral limit around $k=k_0$. This can be used to discriminate different models which give the same spectrum.
\\
\begin{flushright}
YITP-17-25
\end{flushright}

\end{abstract}

\maketitle
\section{Introduction}
Different cosmological observations such as the  Cosmic Microwave Background (CMB) anisotropies and the Large Scale Structures (LSS) have given us observational evidence that the observed universe originated from  fluctuations in the very early universe \cite{Ade:2013zuv,Ade:2015xua,Planck:2013jfk,Ade:2015lrj}. Cosmic inflation, a period of accelerated expansion at early times, is the simplest framework able to explain the origin of these primordial fluctuations and provides a good fit to the data, while alternatives deviating from the inflationary paradigm are less compelling \cite{Ade:2013zuv,Ade:2015xua,Planck:2013jfk,Ade:2015lrj}.
There is a plethora of inflationary models proposed in the literature which can predict the same spectrum of primordial perturbations \cite{Martin:2013tda,2009astro2010S.158K,xc,xc2,Chen:2016vvw,Palma:2014hra}. In this sense deviations from Gaussian statistics of the cosmological density fluctuations, the so-called primordial non-Gaussianities (NG), are important to discriminate between different models~\cite{2009astro2010S.158K,xc,xc2,Byrnes:2015asa,Renaux-Petel:2015bja,Novaes:2013qxa,Novaes:2015uza,Komatsu:2010hc,Chen:2016vvw,Palma:2014hra,Domenech:2016zxn,Chluba:2015bqa,Gariazzo:2015qea,Mooij:2015cxa,Appleby:2015bpw,Hunt:2015iua,Hazra:2016fkm,Gariazzo:2014dla,Romano:2014kla,Cadavid:2015iya,GallegoCadavid:2016wcz,Xu:2016kwz,a2,a3,GallegoCadavid:2017pol}. Recent CMB observations \cite{Ade:2013ydc,Ade:2015ava} have not completely ruled out primordial non-Gaussianity and consequently theoretical predictions could be used in the future to discriminate between different inflationary models.

Observations indicate that the spectrum of primordial curvature perturbations has some deviations from scale invariance and a possible explanation  could be features of the inflaton potential \cite{Leach:2001zf,Starobinsky:1992ts,Adams:2001vc,Chluba:2015bqa,Gariazzo:2015qea,Mooij:2015cxa,Appleby:2015bpw,Hunt:2015iua,Romano:2012kj,Romano:2014kla,Cadavid:2015iya,GallegoCadavid:2016wcz,Hazra:2016fkm,Dorn:2014kga,Gariazzo:2014dla,Nicholson:2009zj,Hunt:2013bha,Chen:2016zuu,Chen:2016vvw,Joy:2008qd,constraints2,Ashoorioon:2006wc,Ashoorioon:2008qr,Ballardini:2016hpi,Pearce:2017bdc,Gong:2017vve} such as a step in the mass or discontinuities of other derivatives~\cite{Cadavid:2015iya}. In the De Sitter limit analytical solutions for the perturbations modes can be derived and used to compute different correlation functions. Using these analytical results we show that there are classes of modified inflaton potentials which produce the same effects on the spectrum and can only be distinguished in certain limits and configurations of the bispectrum.

The paper is organized as follows. In section \ref{m} we introduce the type of modifications of the inflaton potential and give analytical approximations for the background quantities. In section \ref{dpscp} we study the degeneracy of the primordial spectrum of curvature perturbations. In section \ref{bdbl} we show how the degeneracy can be broken at the bispectrum level in certain limits and configurations of the bispectrum. 

\section{Inflation and the model}\label{m}
We consider a single scalar field $\phi$ minimally coupled to gravity according to the action
\begin{equation}\label{action1}
  S = \int d^4x \sqrt{-g} \left[ \frac{1}{2} M^2_{Pl} R  - \frac{1}{2}  g^{\mu \nu} \partial_\mu \phi \partial_\nu \phi -V(\phi)
\right],
\end{equation}
where $R$ is the Ricci scalar, $ M_{Pl} = (8\pi G)^{-1/2}$ is the reduced Planck mass, $g_{\mu \nu}$ is the $FLRW$ metric in a flat universe, and $V$ is the potential energy of the inflaton. The slow roll parameters in terms of conformal time $\tau$ are defined as
\bea \label{slowroll}
  \epsilon \equiv \frac{-H'}{a H^2} \,\,\,\, , \,\,\,\, \eta \equiv \frac{\epsilon'}{a H \epsilon},
\eea
where $a$ is the scale factor, $H$ is the Hubble parameter, and primes indicate derivatives with respect to conformal time.

We will study the degeneracy of the spectrum and bispectrum of primordial perturbations using the following potential \cite{Romano:2014kla} 
\begin{equation}\label{pot}
V(\phi)= \left\{
\begin{array}{lr}
V_{b} + \frac{1}{2}m^2 \phi^2, & \phi \le \phi_0 ,\\
V_{a} + \frac{1}{2}m^2 \phi^2 + \lambda \phi^n, \, & \phi < \phi_0,
\end{array}
\right.
\end{equation}
where $V_{b}$ and $V_{a}$ are  the vacuum energy  before and after the feature respectively, $m$ is the inflaton mass, $\lambda$ is a parameter that controls the magnitude  of the potential modification, and $\phi_0=\phi(\tau_0)$, where $\tau_0$ is the feature time. The condition $V_{a}= V_{b}-\lambda \phi_0^n$ ensures the continuity of the potential at 
$\phi_0$ and the value of $\phi_0$ determines the scale at which the effects of the feature appear in the spectrum and bispectrum of curvature perturbations. In the following sections we will show that the features in the spectrum and bispectrum appear around the scale $k_0=-1/\tau_0$ which 
is leaving the horizon at that time.

\subsection{Analytic approximation for the background equations}
Assuming the De Sitter approximation an analytic approximation for the scalar field before and after the feature was found in \cite{Romano:2014kla}. Before the feature this analytic solution is
\begin{equation}\label{bsb}
  \phi_b(\tau)= \phi_b^+ a(\tau)^{\lambda^+} \, ,
\end{equation}
where $\phi_b^+$ is a constant of integration, $a=(-H\tau)^{-1}$, and
\begin{equation}\label{lambdaplusminus}
  \lambda^{\pm}=\frac{3}{2}\left( -1\pm \sqrt{1- \left(\frac{2 m}{3 H}\right)^2}\right) \,.
\end{equation}
After the feature the analytic solution is
\begin{equation}\label{bsa}%\textsc{e}
  \phi_a(\tau)= \phi_{a}^{{}_{(0)}} +\phi_{a}^{{}_{(1)}} (\tau-\tau_0)+\phi_{a}^{{}_{(2)}} (\tau-\tau_0)^2 + 
  \phi_a^+ a(\tau)^{\lambda^+} + \phi_a^- a(\tau)^{\lambda^-},
\end{equation}
where $\phi_{a}^{{}_{(i)}} (i=0,1,2)$ are constants depending on the parameters $n, \lambda$, and $\phi_0$ \cite{Romano:2014kla}. The $\phi_{a}^{\pm}$ are constants of integration given by
\bea \label{CI}
 \phi_{a}^{\pm} = \frac{\pm1}{a_0{}^{\lambda^{\pm} } (\lambda^--\lambda^+) } \left\{ \lambda^\mp \phi _{0}+ \phi_{0}'\tau_0+\frac{n \lambda 
\phi_{0}{}^{n-2}}{m^2}
	\right. \\ \nonumber
	\left. \times \left[ \lambda^\mp \phi_{0}+ \frac{(n-1)}{(m^2-2 H^2)} \left( (m^2+2 H^2 \lambda^\mp) \phi_{0}'\tau_0 -\lambda^\mp H^2 
\phi_{0}''\tau_0^2 \right)\right] \right\} \, ,
\eea
where quantities evaluated at $\tau_0$ are denoted by the subscript $0$. An analytic 
approximation for the slow roll parameters after the feature is given by \cite{Romano:2014kla}
\bea\label{slowrollapprox}
  \epsilon_a(\tau)\approx \frac{1}{2 M_{Pl}^2}\Biggl(\lambda^+ \phi_a^+ a(\tau)^{\lambda^+}+ \lambda^- \phi_a^- a(\tau)^{\lambda^-} \Biggl)^2 \, , 
\\ \nonumber
  \eta_a(\tau)\approx 2 \frac{ (\lambda^+)^2 \phi_a^+ a(\tau)^{\lambda^+}+ (\lambda^-)^2\phi_a^- a(\tau)^{\lambda^-} }{
		\lambda^+ \phi_a^+ a(\tau)^{\lambda^+} + \lambda^- \phi_a^- a(\tau)^{\lambda^-} }.
\eea

\section{Degeneracy of primordial spectrum of curvature perturbations}\label{dpscp}
In this section we study the case in which the primordial spectrum of curvature perturbations is degenerate at all scales for different values of $n$ and $\lambda$. We adopt the following definition for the spectrum
\begin{equation}\label{ps2}
  P_{\mathcal{R}_{c}}(k) \equiv \frac{k^3}{2\pi^2}|\mathcal{R}_{c}(k)|^2 \,,
\end{equation}
where $k$ is the comoving wave number and $\mathcal{R}_{c}(k)$ is the Fourier transform of the curvature perturbation on comoving slices. We will consider models with the following parameters
\begin{equation}\label{parameters}
  m=6.97\times10^{-9}M_{Pl}, \,\,\,\, H=3.3\times 10^{-7}M_{Pl},\,\,\,\, \phi_b^+=10M_{Pl},
\end{equation}
where we can note that $m\ll H$. From now on we adopt a system of units in which $c=\hbar=M_{Pl}=1$.

In \cite{Romano:2014kla} an analytic approximation for the spectrum of primordial curvature perturbations was derived for the case of the potential in eq.~\eqn{pot} 
\bea\label{mps}
P_{\mathcal{R}_{c}}(k)=\frac{ H^2}{8\pi^2 \epsilon_a(\tau_e)} \left\{1 + \frac{D_0}{k} \left[\left(\frac{k_0^2}{k^2}-1\right)\sin\left(\frac{2 
k}{k_0}\right)-\frac{2k_0}{k} \cos\left(\frac{2 k}{k_0}\right) \right] \right. \nonumber \\ \\ 
	\left. +\frac{D_0^2}{2 k^2} \left[ 1+ \frac{2k_0^2}{k^2} +\frac{k_0^4}{k^4}+\left(1-\frac{k_0^4}{k^4}\right)\cos\left(\frac{2 
k}{k_0}\right)-\frac{2k_0}{k}\left(1+\frac{k_0^2}{k^2}\right)\sin\left(\frac{2 k}{k_0}\right)\right] \right\} \, , \nonumber 
\eea
where $\tau_e$ is the time at the end of inflation. The parameter $D_0$ is related to the discontinuity in $\phi_0''$ produced by the modification of the potential. This implies that the equation for the comoving curvature perturbations
\begin{equation}\label{cpe2}
 u_k''+ \left( k^2 - \frac{z''}{z} \right) u_k =0,
\end{equation}
has a Dirac delta function in $z''/z$, where we defined $ u_k \equiv z \mathcal{R}_{c}(k)$ and $z\equiv a\sqrt{2 \epsilon}$. To evaluate the discontinuity we integrate the Dirac delta function around the feature time 
\cite{Romano:2014kla,pp}
\begin{equation}\label{D0}
  D_0 \equiv \lim_{\delta \to 0} \int_{\tau_0-\delta}^{\tau_0+\delta}\frac{z''}{z}d\tau = -n \lambda a_0^2 \frac{\phi_0^{n-1}}{\phi_0'} \approx n \lambda \phi_0^{n}
\frac{3 k_0}{m^2 \phi_0^2},
\end{equation}
where we have used \cite{Romano:2014kla}
\bea \label{approxs}
 \phi'_0 \approx \lambda^+ k_0 \phi_0, \hspace{0.3cm} a_0=\frac{k_0}{H}, \mbox{ and } \lambda^+ \approx -\frac{1}{3}\frac{m^2}{H^2} . 
\eea

\begin{figure}
 \begin{minipage}{.45\textwidth}
  \includegraphics[scale=0.6]{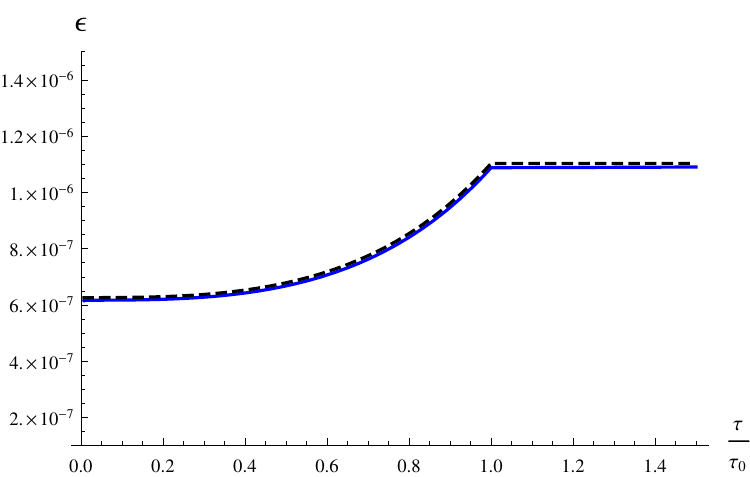}
  \end{minipage}
 \begin{minipage}{.45\textwidth}
  \includegraphics[scale=0.6]{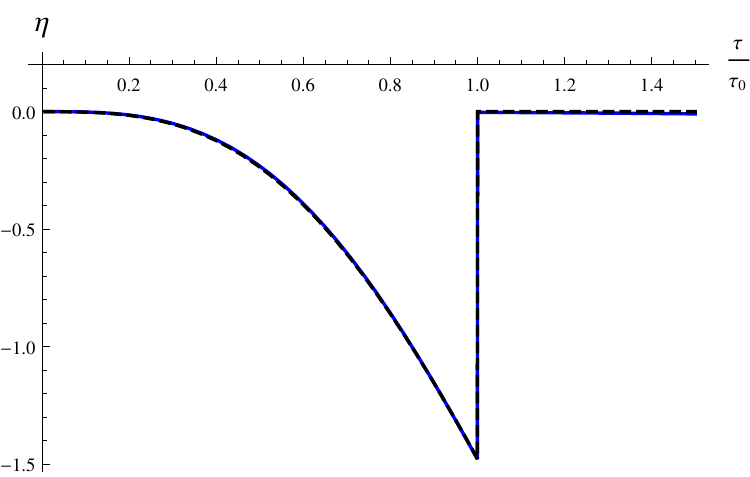}
 \end{minipage}
 \caption{The numerically (blue) and analytically (black-dashed) computed slow-roll parameters are plotted for $n=3$ and $\lambda=-4\times 10^{-19}$. In the  analytic approximation for the slow roll parametes in eq. \eqn{slowrollapprox} we use the  approximation for $\phi_a^{\pm}$ given in eq. \eqn{phiaapprox}.}
\label{srapprox}
\end{figure}

\begin{figure}
 \includegraphics[scale=1]{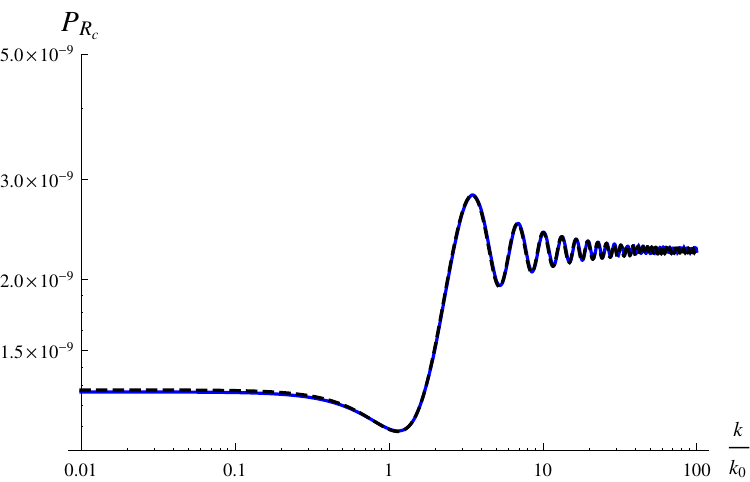}
 \caption{The numerically (blue) and analytically (black-dashed) computed spectrum of curvature perturbations is plotted for $n=3$ and $\lambda=-4\times 10^{-19}$. 
 For the analytic approximation of the spectrum in eq. \eqn{mps} we use the approximation for $\phi_a^{\pm}$ given in eq. \eqn{phiaapprox}.}% for the slow roll parametes in eq. \eqn{slowrollapprox}.}
\label{Pplotapprox}
\end{figure}

\begin{figure}
 \includegraphics[scale=1]{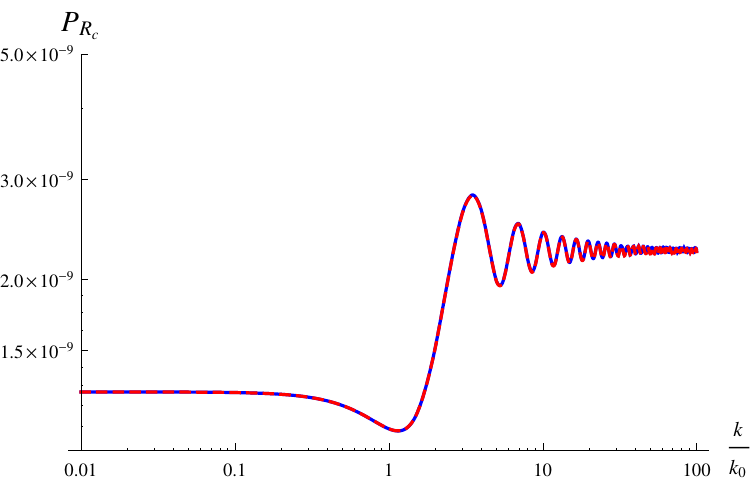}
 \caption{The numerically computed spectrum of curvature perturbations is plotted for $\{n=3, \lambda=-4\times 10^{-19}\}$ (blue) and $\{ n=4, \lambda=-3\times 10^{-20} \}$ (red-dashed), corresponding to $D_0=-0.74$. As it can be seen  the spectrum is the same at all scales, as predicted by the analytical calculation.}
\label{Pplot}
\end{figure}

In order to study the behavior of the spectrum and bispectrum we make a further approximation of eq.~\eqn{CI} and express it in terms of $D_0$ obtaining 
\bea \label{phiaapprox}
\phi_a^+ \approx \phi_0 \Biggl( 1+ \frac{n \lambda \phi_0^{n-2}}{m^2} \Biggr) \approx \phi_0 \Biggl( 1+ \frac{1}{3} \frac{D_0}{k_0} \Biggr) , \\ \nonumber
\phi_a^- \approx - \frac{\lambda^+ }{\lambda^- } \frac{1}{a_0^{\lambda^-}} \frac{n \lambda \phi_0^{n-1}}{m^2} \approx -\phi_0 
\frac{\lambda^+}{\lambda^- } \Biggl(\frac{H}{k_0} \Biggr)^{\lambda^-}\frac{1}{3} \frac{D_0}{k_0} ,
\eea
where we have  used eq. \eqn{approxs} and  \cite{Romano:2014kla}
\bea
\lambda^- \approx -3 \Bigl(1- \frac{m^2}{9H^2}\Bigr) \,, \\
\phi''_0 \approx \lambda^+ k_0^2 \phi_0 \,.
\eea
From now on we will use the new approximation in eq. \eqn{phiaapprox} to derive all the analytic results below. As shown in Fig. \ref{srapprox} and Fig. \ref{Pplotapprox} the new approximation is in good agreement with the numerical results. The above equations are crucial to understand the origin of the degeneracy of the spectrum. From eq. \eqn{slowrollapprox} we can in fact see that the slow roll parameters dependence on $n$ and $\lambda$ is completely determined by $D_0$, and for this reason the spectrum in eq. \eqn{mps} is also only depending on $D_0$. This implies that models with the same $D_0$ but different $n$ and $\lambda$ will have the same spectrum, as long as 
\begin{equation}\label{nlambdaconst}
n \lambda \phi_0^{n} = \mbox{constant}.
\end{equation}
In Fig. \ref{Pplot} we see the degeneracy of the primordial spectrum of curvature perturbations. The results of the spectrum  are plotted using two different sets of  values for $n$ and $\lambda$ corresponding to the same $D_0$. As predicted by the analytic approximation given above, models with the same $D_0$ have the same evolution of the primordial spectrum. 

\section{Breaking of degeneracy with  the bispectrum}\label{bdbl}
As we have seen in the previous section the analytic calculations confirmed by numerical results show that a degeneracy in the spectrum of curvature perturbations is expected for models having the same $D_0$ parameter. In this section we will investigate if this degeneracy is also happening at the bispectrum level.

A common quantity to study the non-Gaussianity is the non-linear parameter $f_{NL}$ defined by 
\bea\label{fNL}
\frac{6}{5} f_{NL}(k_1,k_2,k_3)\equiv 
\frac{B_{\mathcal{R}_{c}}(k_1,k_2,k_3)}{\mathbf{P}_{\mathcal{R}_{c}}(k_1)\mathbf{P}_{\mathcal{R}_{c}}(k_2)+\mathbf{P}_{\mathcal{R}_{c}}(k_1)\mathbf{P}_{\mathcal{R}_{c}}(k_3)+\mathbf{P}_{\mathcal{R}_{c}}
(k_2)\mathbf{P}_{\mathcal{R}_{c}}(k_3)} \, ,
\eea
where $B_{\mathcal{R}_{c}}$ is the bispectrum of primordial curvature perturbations given by
\bea \label{b}
  B_{\mathcal{R}_{c}}(k_1,k_2,k_3)= 2 \Im\Biggl[ \mathcal{R}_{c}(k_1,\tau_e) \mathcal{R}_{c}(k_2,\tau_e)\mathcal{R}_{c}(k_3,\tau_e) \int^{\tau_e}_{\tau_0} d\tau  \eta(\tau) \epsilon(\tau) a(\tau)^2 \\ \nonumber
  \Biggl( 2\mathcal{R}_{c}^*(k_1,\tau)\mathcal{R}_{c}^{'*}(k_2,\tau)\mathcal{R}_{c}^{'*}(k_3,\tau) - k^2_1 \mathcal{R}_{c}^*(k_1,\tau)\mathcal{R}_{c}^*(k_2,\tau)\mathcal{R}_{c}^*(k_3,\tau) \Biggr) \\ \nonumber 
  + \mbox{ two permutations of } k_1, k_2,\mbox{ and } k_3  \Biggr] \, ,
\eea
and
\begin{equation}
\mathbf{P}_{\mathcal{R}_{c}} \equiv \frac{2\pi^2}{k^3} P_{\mathcal{R}_{c}} \, .
\end{equation}
If we replace $\mathbf{P}_{\mathcal{R}_{c}}$ in eq. (\ref{fNL}) we obtain  $f_{NL}$ in terms of our definition of $P_{\mathcal{R}_{c}}$
\bea
f_{NL}(k_1,k_2,k_3)= \frac{10}{3}\frac{(k_1 k_2 k_3)^3}{(2\pi)^4} 
\frac{B_{\mathcal{R}_{c}}(k_1,k_2,k_3)}{P_{\mathcal{R}_{c}}(k_1)P_{\mathcal{R}_{c}}(k_2)k^3_3+P_{\mathcal{R}_{c}}(k_1)P_{\mathcal{R}_{c}}(k_3)k^3_2+P_{\mathcal{R}_{c}}(k_2)P_{\mathcal{R}_{c}}(k_3)k^3_1} \, .
\eea
In this paper we will use the following quantity to study the degeneracy of the primordial bispectrum \cite{Romano:2014kla,Cadavid:2015iya}
\begin{equation}\label{FNL}
  F_{NL}(k_1,k_2,k_3;k_*)\equiv \frac{10}{3(2\pi)^4}\frac{(k_1 k_2 k_3)^3}{k_1^3+k_2^3+k_3^3}\frac{B_{\mathcal{R}_{c}}(k_1,k_2,k_3)}{P_{\mathcal{R}_{c}}^2(k_*)} \, ,
\end{equation}
{where $k_*$ is the pivot scale at which the power spectrum is normalized, i.e., $P_{\mathcal{R}_{c}}(k_*)\approx 2.2\times 10^{-9}$. In the equilateral limit 
our definition of $F_{NL}$ reduces to $f_{NL}$ if the spectrum is approximately scale invariant, but in general $f_{NL}$ and $F_{NL}$ are different. 
In the squeezed limit for instance  they are not the same, but $F_{NL}$ still provides useful information about the non-Gaussian behavior of 
$B_{\mathcal{R}_{c}}$ although they cannot be compared directly.

\subsection{Analytic approximation for the Bispectrum}

\subsection*{Squeezed and equilateral limits at large scales}
In the large scale isosceles configuration $k_2=k_3=k\ll k_0$ eq. \eqn{FNL} reduces to the following analytic formula 
\cite{Romano:2014kla}
\bea\label{fnlb}
F_{NL}^{<}(k_1,k) \approx -\frac{5}{6}\frac{H^5}{(2\pi)^4P_{\mathcal{R}_{c}}^2(k_*)} \frac{a(\tau_e)}{(\lambda^+)^3 (\phi_b^+)^6} \frac{\phi_a^{+2}}{k} 
\Biggl[\frac{2k+k_1}{k_0}\cos{\left(\frac{2k+k_1}{k_0}\right)}\\ \nonumber
+\left(\frac{k}{k_0}\frac{2k_1+k}{k_0}-1\right)\sin{\left(\frac{2k+k_1}{k_0}\right)} \Biggr]\,.
\eea
In the squeezed limit $k_1\ll k$ we get
\bea\label{fnlbsl}
F_{NL}^{< \, SL}(k) \approx -\frac{5}{6}\frac{H^5}{(2\pi)^4P_{\mathcal{R}_{c}}^2(k_*)} \frac{a(\tau_e)}{(\lambda^+)^3 (\phi_b^+)^6} \frac{\phi_a^{+2}}{k} 
\Biggl[\frac{2k}{k_0}\cos{\left(\frac{2k}{k_0}\right)}+\left(\frac{k^2}{k_0^2}-1\right)\sin{\left(\frac{2k}{k_0}\right)} \Biggr]\,,
\eea
and  in the equilateral limit $k_1= k$
\bea\label{fnlbel}
F_{NL}^{< \, EL}(k) \approx -\frac{5}{6}\frac{H^5}{(2\pi)^4P_{\mathcal{R}_{c}}^2(k_*)} \frac{a(\tau_e)}{(\lambda^+)^3 (\phi_b^+)^6} \frac{\phi_a^{+2}}{k} 
\Biggl[\frac{3k}{k_0}\cos{\left(\frac{3k}{k_0}\right)}+\left(\frac{3k^2}{k_0^2}-1\right)\sin{\left(\frac{3k}{k_0}\right)} \Biggr]\,.
\eea
The results of the numerical and analytic approximation of the bispectrum are shown in Fig.~\ref{plotfnlb} and are in good agreement far from $k_0$ both in the squeezed and equilateral limits.
\begin{figure}
 \begin{minipage}{.45\textwidth}
  \includegraphics[scale=0.6]{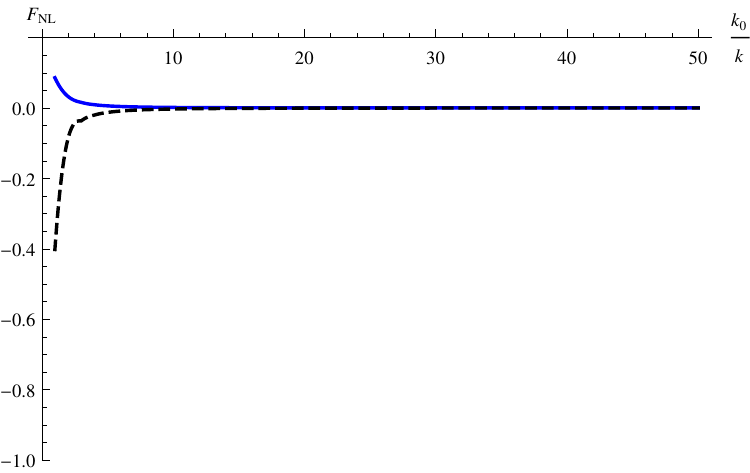}
  \end{minipage}
  \begin{minipage}{.45\textwidth}
  \includegraphics[scale=0.6]{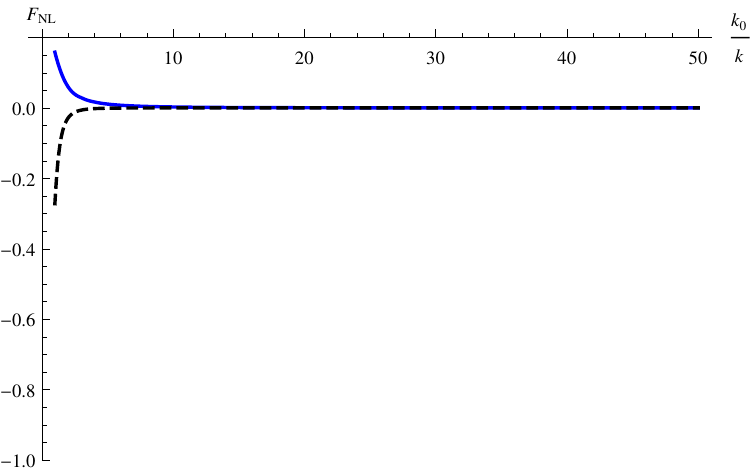}
  \end{minipage}
 \caption{The numerically (blue) and analytically (black-dashed) computed large scales $F_{NL}$ in the squeezed (left) and equilateral (right) limit is plotted for $n=3$ and $\lambda=-4\times 10^{-19}$.}
\label{plotfnlb}
\end{figure}
\subsection*{Squeezed and equilateral limits at small scales}
In the small scale isosceles configuration, when $k_2=k_3 \gg k_0$ and $k_1\gg k_0$ a fully analytic template is given by \cite{Romano:2014kla} 
\bea \label{fnla}
F_{NL}^>(k,k_2)\approx \frac{20}{3(2\pi)^4}\frac{(k k_2^2)^3}{(k^3+2k_2^3)P_{\mathcal{R}_{c}}^2(k_*)} \Im \biggl[ \mathcal{R}_{c}(k,\tau_e) \mathcal{R}_{c}(k_2,\tau_e)^2 \biggl( 4 I_1(k_2,k,k_2) \\ \nonumber
+ 2 I_1(k,k_2,k_2) - (k^2+2k_2^2) I_2(k,k_2,k_2)  \biggr) \biggr] \, ,
\eea
where $k= k_1+\delta$ and $\delta$ is a phase shift parameter which varies for different models or limits, and
\bea \label{integral12}
I_i (k_1,k_2,k_3) \approx \lambda^+ (\lambda^-)^2  \phi_a^+ \phi_a^- \mathcal{A}_i(\tau_0,k_1,k_2,k_3,q_1) 
+ (\lambda^-)^3 (\phi_a^-)^2 \mathcal{A}_i(\tau_0,k_1,k_2,k_3,q_2) ,
\eea
with $i=1,2$, $q_1=2+\lambda^-$, and $q_2 =2+2\lambda^-$. The $\mathcal{A}_i$ functions are written in the Appendix \ref{A1}. The analytic approximation for the curvature perturbation mode $\mathcal{R}_{c}$ after the feature is given by \cite{Starobinsky:1992ts,Starobinsky:1998mj,Romano:2014kla}
\begin{equation}\label{r}
  \mathcal{R}_{c}(k,\tau)=\frac{1}{ a(\tau)\sqrt{2\epsilon_a(\tau)} }\left[\alpha_k v_k(\tau)+\beta_k v_k^*(\tau)\right] \,,
\end{equation}
where
\begin{equation}\label{alphabeta}
  \alpha_k=1+ \mathrm{i} D_0 |v_k(\tau_0)|^2 \: \mbox{ and } \: \beta_k= -\mathrm{i} D_0 v_k(\tau_0)^2
\end{equation}
are the Bogoliubov coefficients and $v$ is the Bunch-Davies vacuum. 

In the small scale squeezed limit, when $k_2=k_3 \gg k_1>k_0$ eq. \eqn{fnla} reduces to
\bea \label{fnlasl}
F_{NL}^{> \, SL}(k,k_2)\approx \frac{20}{3(2\pi)^4}\frac{(k k_2)^3}{P_{\mathcal{R}_{c}}^2(k_*)} \Im \biggl[ \mathcal{R}_{c}(k,\tau_e) \mathcal{R}_{c}(k_2,\tau_e)^2 \biggl( I_1(k,k_2,k_2) \\ \nonumber
  + 2 I_1(k_2,k,k_2)-k_2^2 I_2(k,k_2,k_2)  \biggr) \biggr] \, ,
\eea
and in the equilateral limit $k_1=k_2=k_3\equiv k$ 
\bea \label{fnlael}
  F_{NL}^{> \, EL}(k)\approx \frac{20}{3(2\pi)^4}\frac{k^6}{P_{\mathcal{R}_{c}}^2(k_*)}\Im \biggl[ \mathcal{R}_{c}(k,\tau_e)^3 \Bigl( 2 I_1(k) -k^2 I_2(k) \Bigr) \biggr] \, .
\eea
The numerical results of the bispectrum and the analytic template eq. \eqn{fnlasl} and eq. \eqn{fnlael} are in good agreement far from $k_0$ both in the squeezed and equilateral limits as shown in Fig.~\ref{plotfnla}. In Fig. \ref{FNLfnlel} we show (in the particular case of the equilateral limit) that our analytic approximation is in good agreement far from $k_0$ but not around the feature scale.
\begin{figure}
 \begin{minipage}{.45\textwidth}
  \includegraphics[scale=0.6]{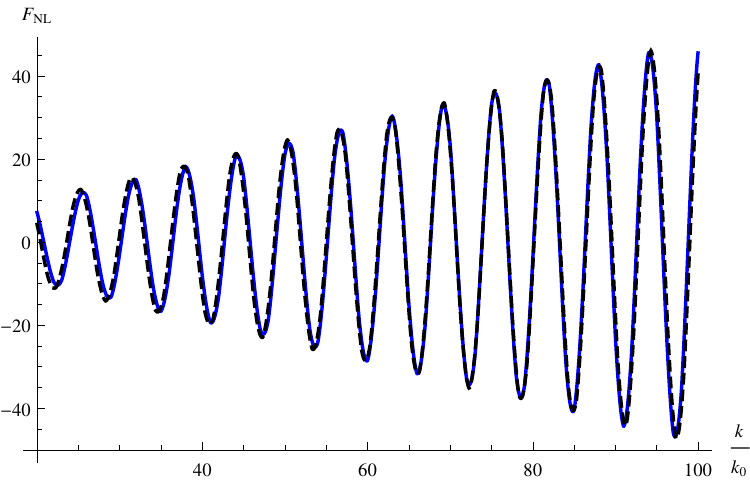}
  \end{minipage}
 \begin{minipage}{.45\textwidth}
  \includegraphics[scale=0.6]{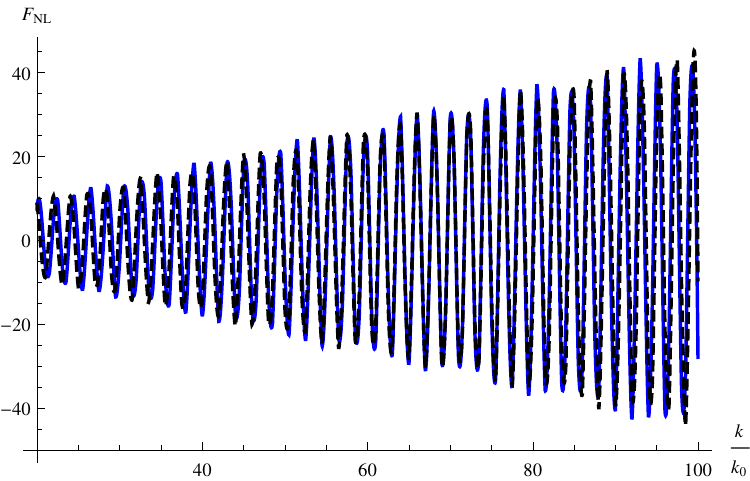}
 \end{minipage}
 \caption{The numerically (blue) and analytically (black-dashed) computed small scales $F_{NL}$ in the squeezed (left) and equilateral (right) limit is plotted for $n=3$ and $\lambda=-4\times 10^{-19}$. In the analytic approximation we use $\delta=2.7k_0$ and $\delta=0.5k_0$ in the squeezed and equilateral limits respectively.}
\label{plotfnla}
\end{figure}

\begin{figure}
 \includegraphics[scale=1]{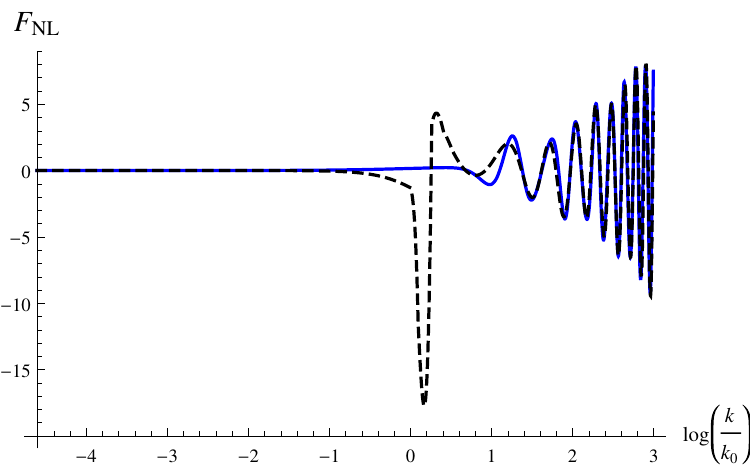}
 \caption{The numerically (blue) and analytically (black-dashed) computed $F_{NL}$ in the equilateral limit is plotted for large and small scales around $k_0$ for $n=3$ and $\lambda=-4\times 10^{-19}$. As can be seen the analytic approximation is good for large and small scales but not around $k_0$.}
\label{FNLfnlel}
\end{figure}
\subsection{Degeneracy of the bispectrum far from $k_0$}
The formulas in eq. \eqn{fnlb} and eq. \eqn{fnla} are essential to understand the degeneracy of the bispectrum because they show that $F_{NL}$ depends on $n$ and $\lambda$ only through the parameter $D_0(n,\lambda)$, implying a degeneracy when eq. \eqn{nlambdaconst} is satisfied, as long as the analytical approximation is valid.

In the large scale case it is easy to see from eq. \eqn{fnlb} that the bispectrum depends on $n$ and $\lambda$ only through $D_0$ since $\phi_a^+$ and $P_{\mathcal{R}_{c}}$ are completely determined by $D_0$, as we saw in previous sections. Thus the bispectrum can be degenerate at large scales in the squeezed and equilateral limits.

In the small scale case we can see from eq. \eqn{fnla} that the bispectrum depends on the spectrum $P_{\mathcal{R}_{c}}$, the curvature perturbation $\mathcal{R}_{c}$ after the feature, and the integrals $I_i$ ($i=1,2$). We already know that the spectrum is completely determined by $D_0$ while from eq. \eqn{r} and eq. \eqn{alphabeta} we can see that $\mathcal{R}_{c}$ depends on $n$ and $\lambda$ only through $D_0$. 
As for the integrals $I_i$ defined in eq. \eqn{integral12} we can see that they depend on $\phi_a^{\pm}$ which are determined by $D_0$ and on the $\mathcal{A}_i$ functions which depend on $D_0$ through the Bogoliubov coefficients defined in eq. \eqn{alphabeta}. Thus again the bispectrum can be degenerate at small scales in the squeezed and equilateral limits.

In Fig. \ref{FNLb} and Fig. \ref{FNLa} we show the degeneracy of the bispectrum of primordial curvature perturbations at large and small scales respectively when $D_0$ has the same value for different choices of $n$ and $\lambda$. 
The dependence of analytical expressions for the large and small scale $F_{NL}$ on $n$ and $\lambda$ only through $D_0$  explains why there is a degeneracy of the bispectrum on scales far from $k_0$. As shown in different figures the numerical calculations confirm the existence of this degeneracy predicted by the analytical calculations.

\begin{figure}
 \begin{minipage}{.45\textwidth}
  \includegraphics[scale=0.6]{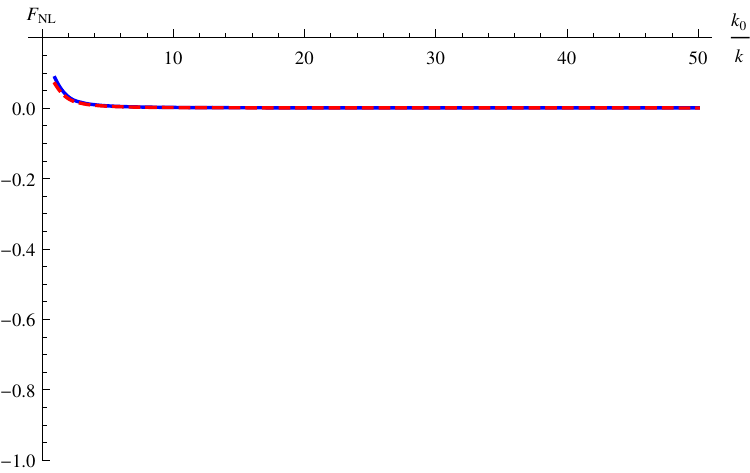}
  \end{minipage}
  \begin{minipage}{.45\textwidth}
  \includegraphics[scale=0.6]{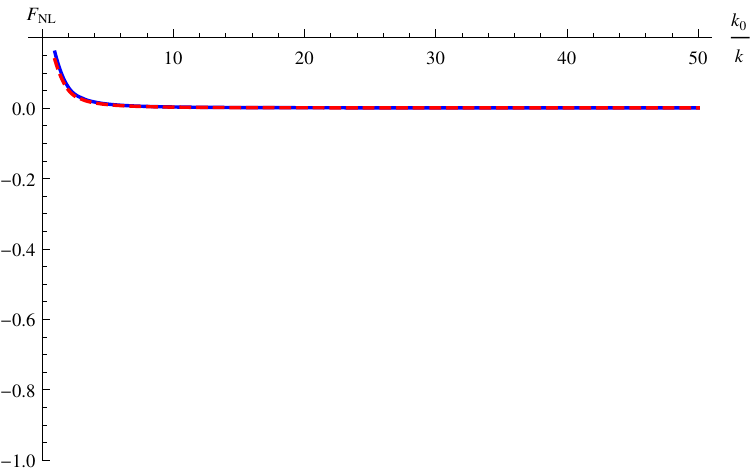}
  \end{minipage}
 \caption{The numerically computed large scales $F_{NL}$ in the squeezed (left) and equilateral (right) limit  is plotted for $n$ and $\lambda$ in the case for which  $D_0$ is the same. The parameters are  $\{n=3, \lambda=-4\times 10^{-19}\}$ (blue) and $\{ n=4, \lambda=-3\times 10^{-20} \}$ (red-dashed), corresponding to the same $D_0=-0.74$.}
\label{FNLb}
\end{figure}

\begin{figure}
 \begin{minipage}{.45\textwidth}
  \includegraphics[scale=0.6]{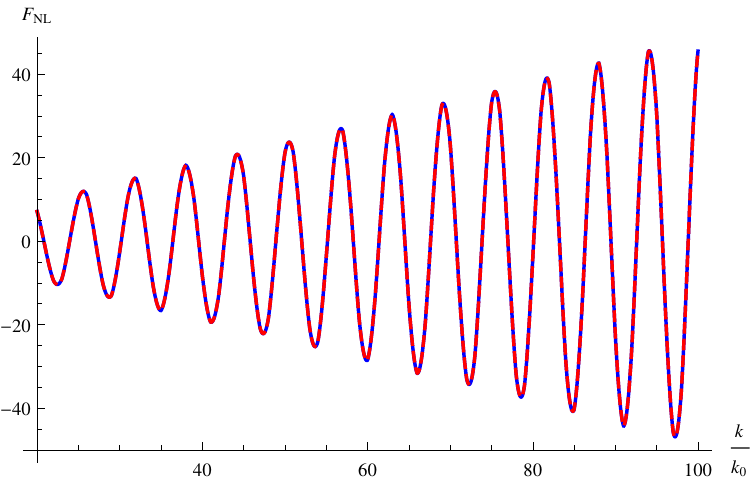}
  \end{minipage}
 \begin{minipage}{.45\textwidth}
  \includegraphics[scale=0.6]{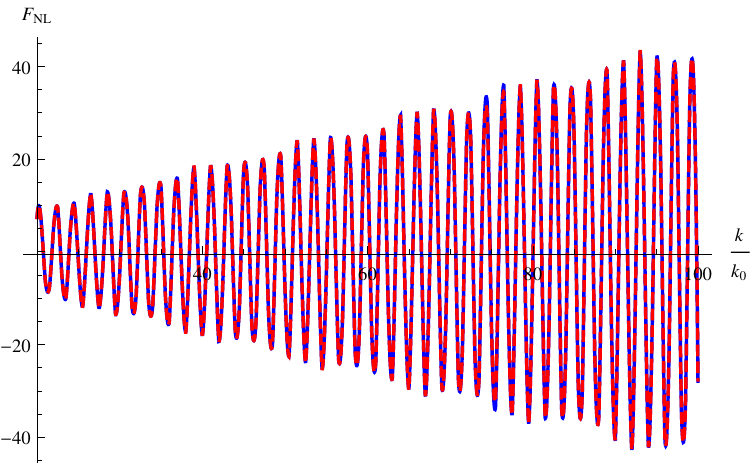}
 \end{minipage}
 \caption{The numerically computed small scales $F_{NL}$ in the squeezed (left) and equilateral (right) limit  is plotted for $\{n=3,\lambda=-4\times 10^{-19}\}$ (blue) and $\{n=4,\lambda=-3\times 10^{-20}\}$ 
(red-dashed), corresponding to $D_0=-0.74$.}
\label{FNLa}
\end{figure}

\subsection{Breaking of degeneracy in the equilateral limit around $k_0$}
Around $k_0$ the analytical approximations used to compute the bispectrum may  not be as accurate as on small and large scales, and consequently a numerical calculation is necessary. In Fig. \ref{FNLbreaking} we show, for different models with the same $D_0$, the numerically computed bispectrum for large and small scales in the squeezed and equilateral limits. The degeneracy is broken only in the equilateral limit around $k_0$, while the squeezed limit is degenerate on any scale. For  larger values of $D_0$  the breaking of the degeneracy is more evident as shown at the bottom of Fig. \ref{FNLbreaking}.

The breaking of the degeneracy of the bispectrum can be explained from the fact that in the integral to compute $F_{NL}$, the integrand depends on $a^2 \epsilon \eta$, which is different for the two models since $\eta$ is different, while $\epsilon$ is approximately the same, as shown in Fig.~(\ref{fig:sr}).

\begin{figure}
 \begin{minipage}{.45\textwidth}
  \includegraphics[scale=0.6]{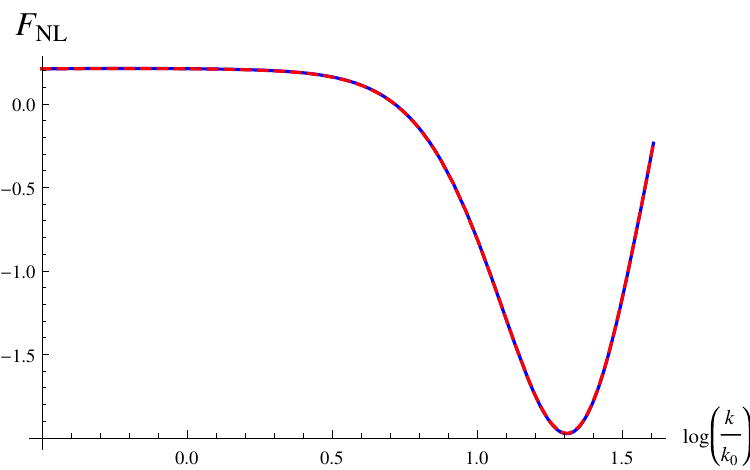}
  \end{minipage}
  \begin{minipage}{.45\textwidth}
  \includegraphics[scale=0.6]{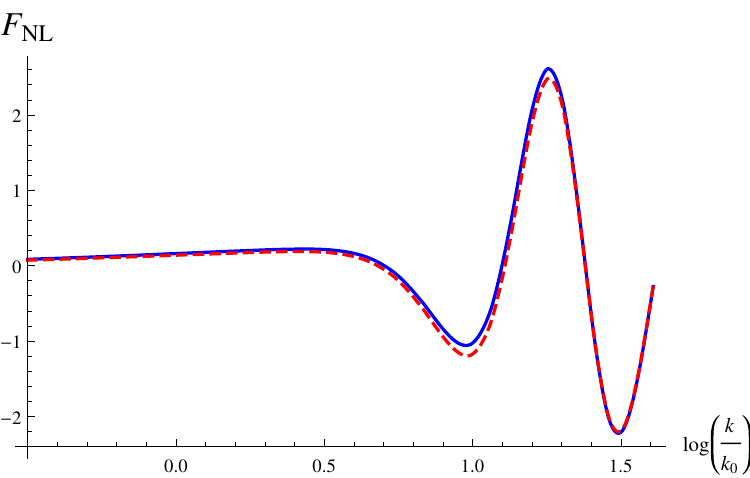}
  \end{minipage}
 \begin{minipage}{.45\textwidth}
  \includegraphics[scale=0.6]{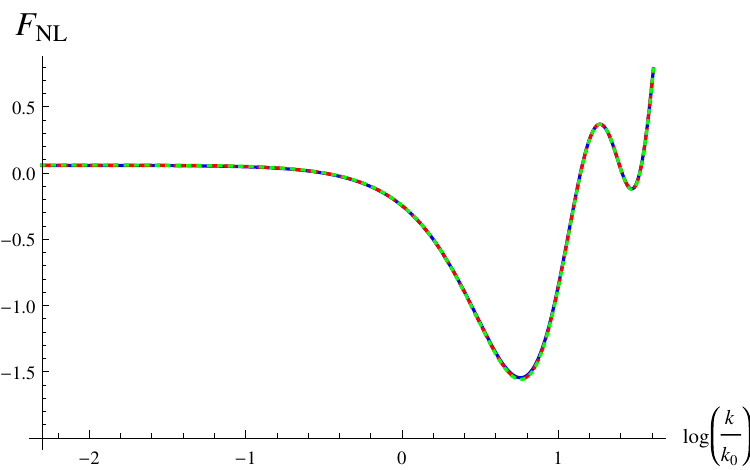}
  \end{minipage}
 \begin{minipage}{.45\textwidth}
  \includegraphics[scale=0.6]{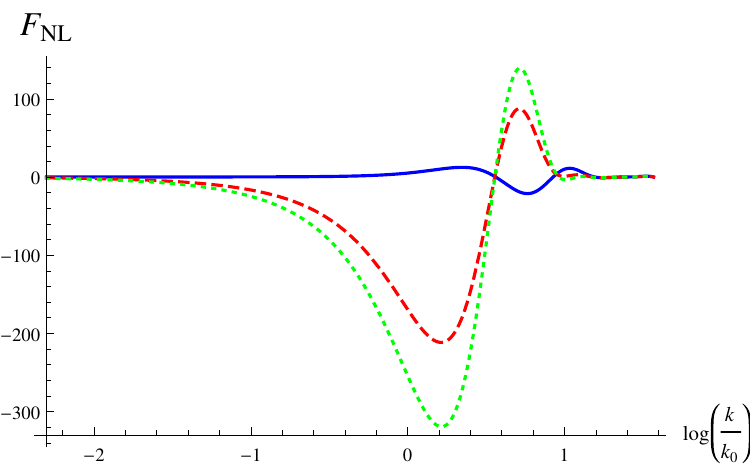}
 \end{minipage}
 \caption{The numerically computed $F_{NL}$ in the squeezed (left) and equilateral (right) limits for large and small scales around $k_0$ is plotted for $n$ and $\lambda$ for models that have the same $D_0$. On the top we have chosen $\{n=3, \lambda=-4\times 10^{-19}\}$ (blue) and $\{ n=4, \lambda=-3\times 10^{-20} \}$ (red-dashed) with $D_0=-0.74$ in both cases. 
On the bottom we have chosen a larger value of $D_0=-6$ in order to see a larger breaking of the degeneracy; these values correspond to $\{ n=2/3, \lambda=-2.3\times 10^{-15} \}$ (blue), $\ n=3, \lambda=-2.4\times 10^{-18} \}$ (red-dashed), and $ \{ n=4, \lambda=-1.8\times 10^{-19}\}$ (green-dotted).}
\label{FNLbreaking}
\end{figure}

\begin{figure}
 \begin{minipage}{.45\textwidth}
  \includegraphics[scale=0.6]{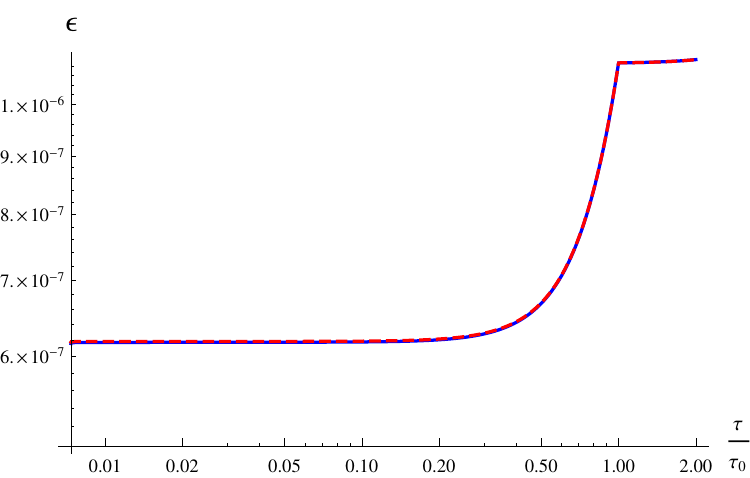}
  \end{minipage}
 \begin{minipage}{.45\textwidth}
  \includegraphics[scale=0.6]{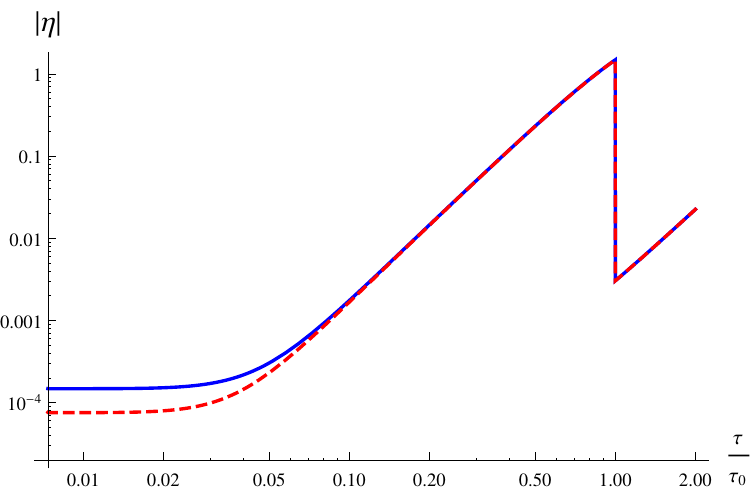}
 \end{minipage}
 \caption{The numerically computed slow-roll parameters $\epsilon$ (left) and $\eta$ (right) are plotted using the potential in Eq.~\eqn{pot} for $n=3$ and $\lambda=-4\times 10^{-19}$ (blue) and $n=4$ and $\lambda=-3\times 10^{-20}$ 
(red-dashed), corresponding to $D_0=-0.74$.}
\label{fig:sr}
\end{figure}

\begin{figure}
  \includegraphics[scale=1]{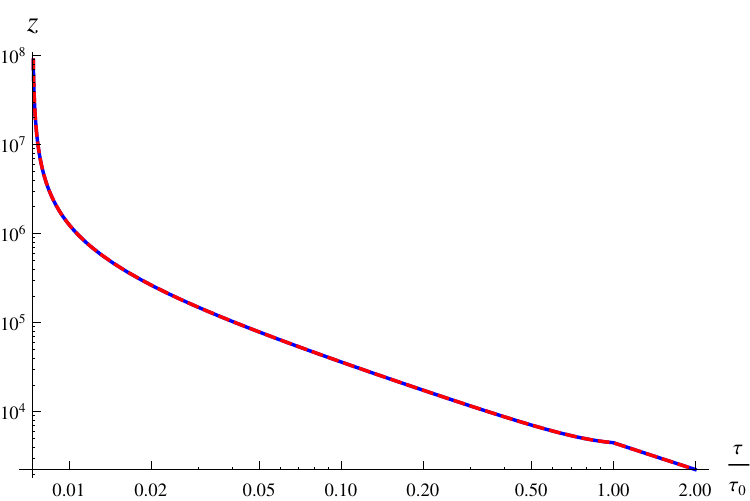}
 \caption{The numerical calculation of $z$  is plotted using the potential in Eq.~\eqn{pot} and for $n$ and $\lambda$ in the case for which  $D_0$ is 
the same. The parameters are  $n=3$ and $\lambda=-4\times 10^{-19}$ (blue) and $n=4$ and $\lambda=-3\times 10^{-20}$ (red-dashed), corresponding to the same $D_0=-0.74$.}
\label{fig:z}
\end{figure}

\section{Consistency relations and  smooth potentials}
It can be shown~\cite{osra} that there is an infinite set of slow-roll parameters histories which can produce the same spectrum of curvature perturbations. This implies that there is no general one-to-one correspondence between the spectrum and higher order correlation functions. In Fig.~\ref{fig:z} we show how different values of the model parameters can give the same $z$, leading to the same predictions of the spectrum since it is the relevant quantity in the calculation of the comoving curvature perturbations $\mathcal{R}_c$ (see Eq.~\eqn{cpe2}). Even in this case, the evolution of the slow-roll parameters might be different, as in the case of $\eta$ as can be seen from Fig.~\ref{fig:sr}. This might lead to different predictions of the bispectrum (see Eq.~\eqn{b}). This freedom implies that in general there can be models like the ones we have studied violating the consistency relations derived for example in Refs.~\cite{Palma:2014hra, Chung:2005hn, Mooij:2015cxa}. These  relations are in fact based on the  approximation  
\be\label{zppzapprox}
\frac{z''}{z} \approx  2a^2H^2 (1 -\frac{1}{4}\tau \eta' )
\ee
which is not accurate for our models, as shown in Fig~\ref{zppz}.

\begin{figure}
\includegraphics[scale=1]{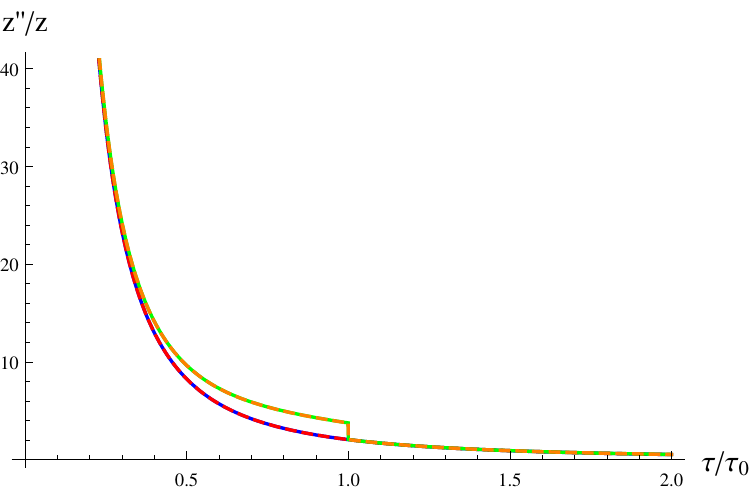}
\caption{For  the potential in Eq.~\eqn{pot} the numerically computed $z''/z$ (blue and red lines)   and the approximation in Eq.~\eqn{zppzapprox} (green and orange lines) are plotted for $n$ and $\lambda$ in the cases for which  
$D_0$ is the same. The parameters are  $n=3$ and $\lambda=-4\times 10^{-19}$ (blue and green) and $n=4$ and $\lambda=-3\times 10^{-20}$ (red-dashed and orange), corresponding to the same $D_0=-0.74$. The approximation is not accurate around the feature. }
\label{zppz}
\end{figure}

\begin{figure}
 \begin{minipage}{.45\textwidth}
  \includegraphics[scale=0.6]{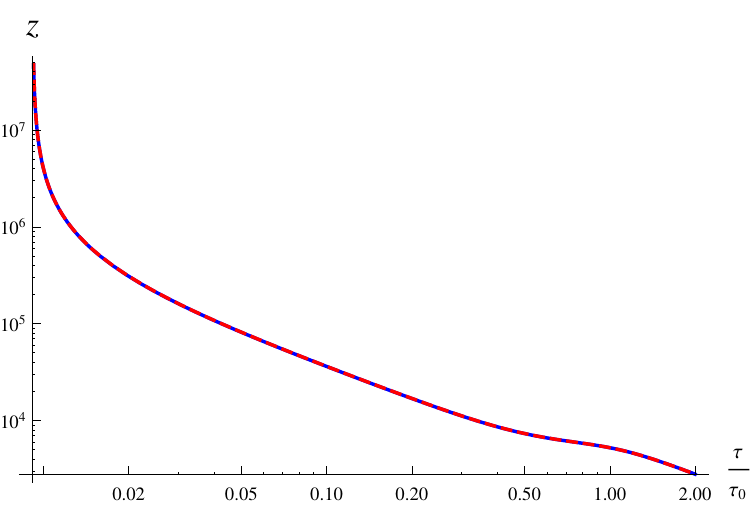}
  \end{minipage}
 \begin{minipage}{.45\textwidth}
  \includegraphics[scale=0.6]{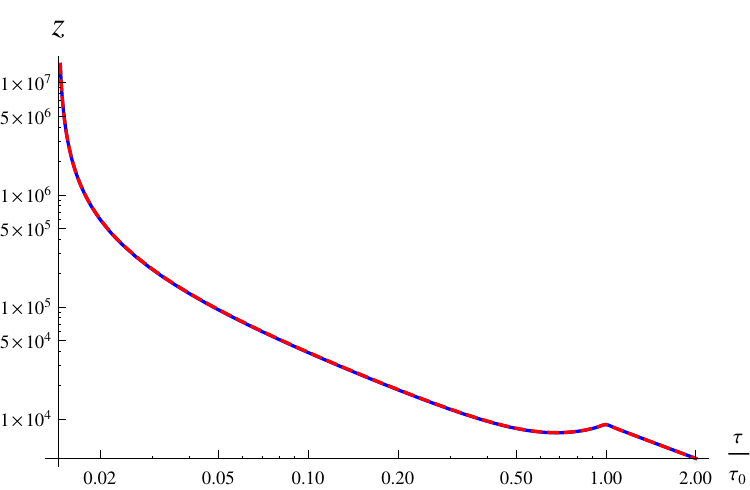}
  \end{minipage}
\begin{minipage}{.45\textwidth}
  \includegraphics[scale=0.6]{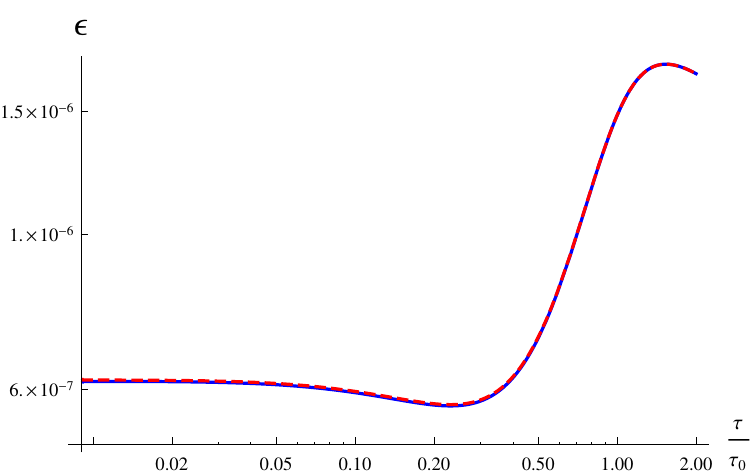}
  \end{minipage}
 \begin{minipage}{.45\textwidth}
  \includegraphics[scale=0.6]{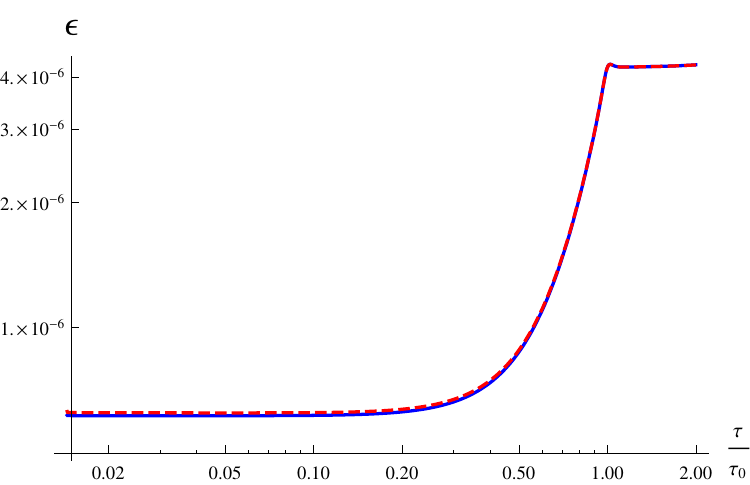}
  \end{minipage}
\begin{minipage}{.45\textwidth}
  \includegraphics[scale=0.6]{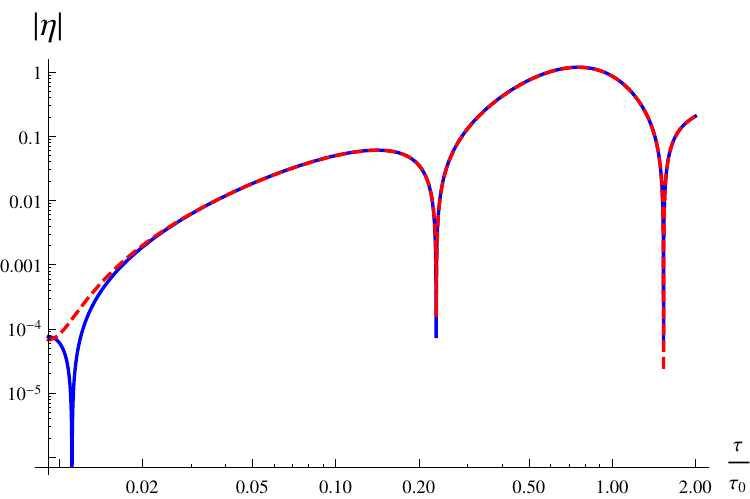}
  \end{minipage}
 \begin{minipage}{.45\textwidth}
  \includegraphics[scale=0.6]{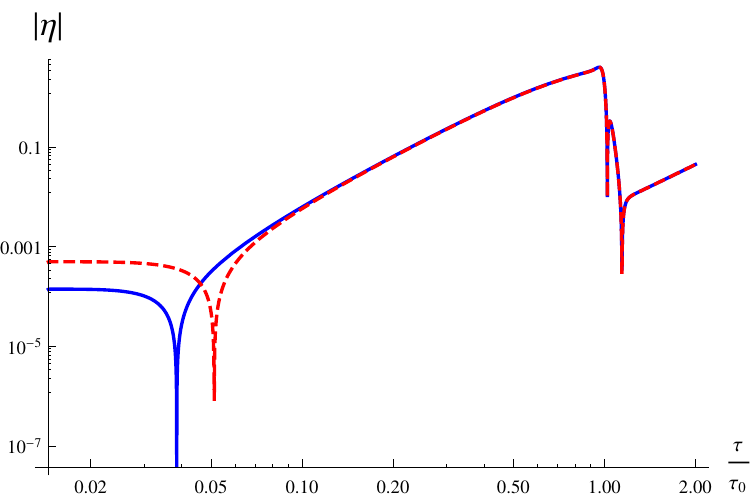}
  \end{minipage}
 \caption{The numerically computed $z, \epsilon$ and $\eta$ are plotted using the potential in Eq.~\eqn{pottanh} for $n$ and $\lambda$ in the cases for which the spectra are the same. On the left we use the parameters $\sigma=10^{-3}$ and $n=3$, $\lambda=-8\times 10^{-19}$ (blue) and  $n=4$, $\lambda=-6\times 10^{-20}$ (red). On the right we use $\sigma=10^{-4}$ and $n=3$, $\lambda=-2.0\times 10^{-18}$ (blue) and  $n=4$, $\lambda=-1.5\times 10^{-19}$ (red).}
\label{zepsetamooth}
\end{figure}

\begin{figure}
 \begin{minipage}{.45\textwidth}
  \includegraphics[scale=0.6]{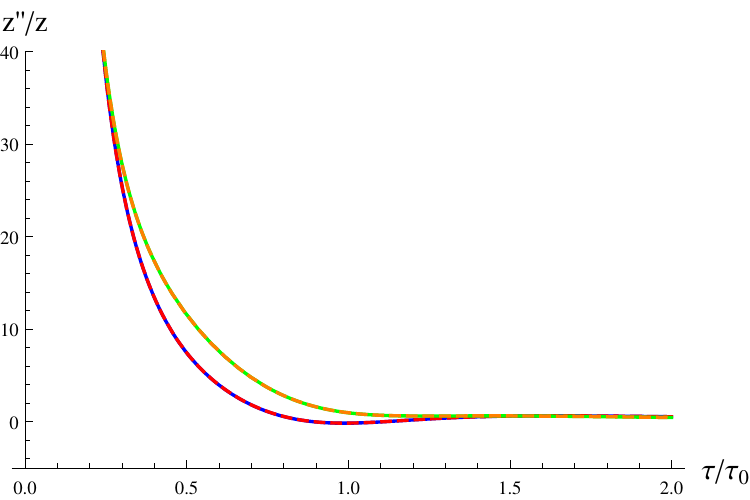}
  \end{minipage}
 \begin{minipage}{.45\textwidth}
  \includegraphics[scale=0.6]{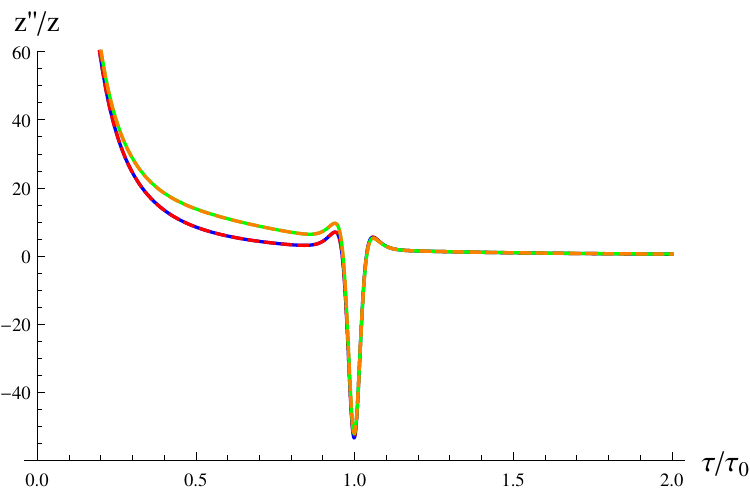}
  \end{minipage}
 \caption{The numerically computed $z''/z$ is plotted using the potential in Eq.~\eqn{pottanh} and the exact numerical result (blue and red lines) and the approximation in Eq.~\eqn{zppzapprox} (green and orange lines) for $n$ and $\lambda$ in the case for which the spectrum is the same. On the left we use the parameters $\sigma=10^{-3}$ and $n=3$, $\lambda=-8\times 10^{-19}$ (blue) and  $n=4$, $\lambda=-6\times 10^{-20}$ (red). On the right we use $\sigma=10^{-4}$ and $n=3$, $\lambda=-2.0\times 10^{-18}$ (blue) and  $n=4$, $\lambda=-1.5\times 10^{-19}$ (red).}
\label{zppzsmooth}
\end{figure}

In order to show that the breaking of the degeneracy in the equilateral limit is not an artifact due to the non-smoothness of the potential we now consider the continuous potential
\begin{equation}\label{pottanh}
V(\phi)= V_0 + \frac{1}{2}m^2 \phi^2 + \frac{1}{2} \lambda (\phi^n-\phi_0^n) \Bigl[1+ \tanh{(\frac{\phi-\phi_0}{\sigma})} 
\Bigr],
\end{equation}
which is equivalent to the potential considered before in Eq.~\eqn{pot} in the limit $\sigma \to 0$. In Fig.~\ref{zepsetamooth} we show that there are  models for which $z$ and $\epsilon$ are approximately the same but $\eta$ is different. In Fig.~\ref{zppzsmooth} we show that the approximation in Eq.~\eqn{zppzapprox} is not accurate around the feature time.

In Figs.~\ref{Pplotstanh} and~\ref{FNLtogether} we show $P_{\mathcal{R}_c}$ and $F_{NL}$ in the squeezed and equilateral limits for the continuous potential. As can be seen from the plots we obtain  results similar to those obtained with the discontinuous potential, namely, the primordial spectrum is degenerate and the degeneracy is only broken in the equilateral limit around $k_0$. We also obtain that the degeneracy is larger for a larger $D_0$ or a steeper transition as in the previous case of a discontinuous potential.

\begin{figure}
 \begin{minipage}{.45\textwidth}
  \includegraphics[scale=0.6]{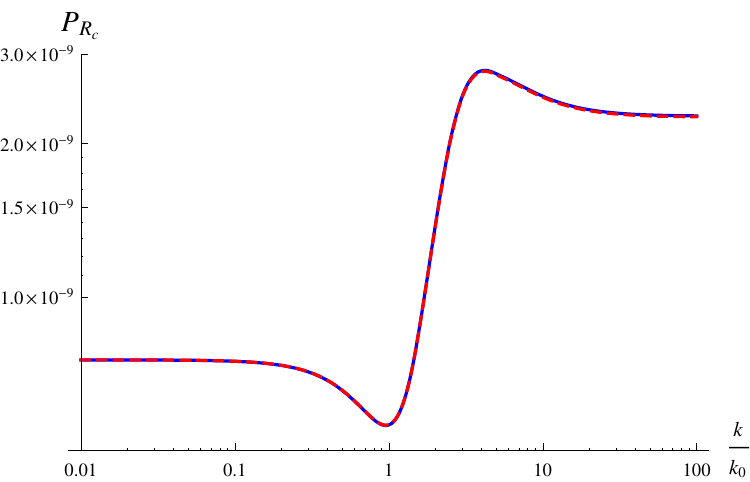}
  \end{minipage}
 \begin{minipage}{.45\textwidth}
  \includegraphics[scale=0.6]{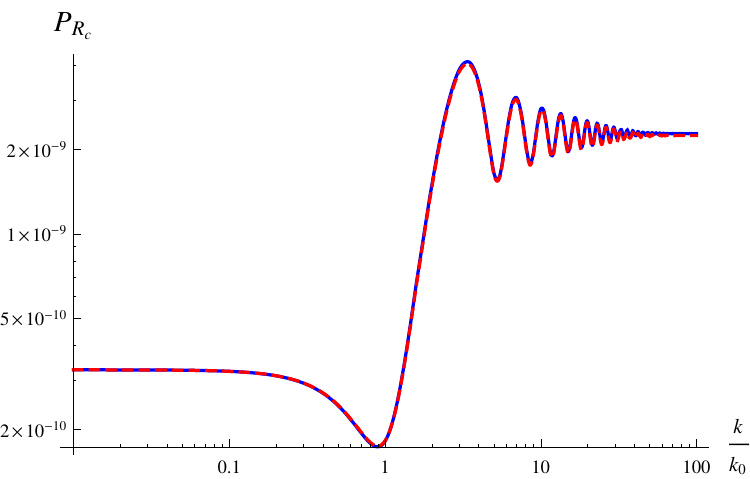}
  \end{minipage}
 \caption{The numerically computed spectrum of primordial curvature perturbations $P_{\mathcal{R}_c}$ is plotted using the potential in Eq.~\eqn{pottanh}. On the left we use the parameters $\sigma=10^{-3}$ and $n=3$, $\lambda=-8\times 10^{-19}$ (blue) and  $n=4$, $\lambda=-6\times 10^{-20}$ (red). On the right we use $\sigma=10^{-4}$ and $n=3$, $\lambda=-2.0\times 10^{-18}$ (blue) and  $n=4$, $\lambda=-1.5\times 10^{-19}$ (red).}
\label{Pplotstanh}
\end{figure}

\begin{figure}
 \begin{minipage}{.45\textwidth}
  \includegraphics[scale=0.6]{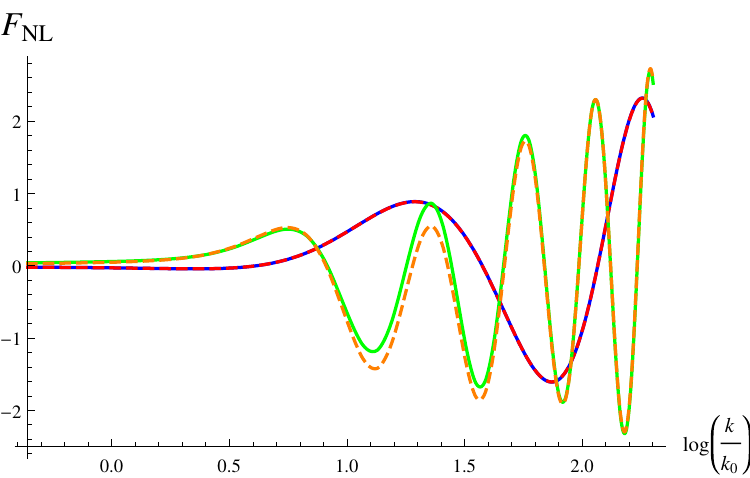}
  \end{minipage}
 \begin{minipage}{.45\textwidth}
  \includegraphics[scale=0.6]{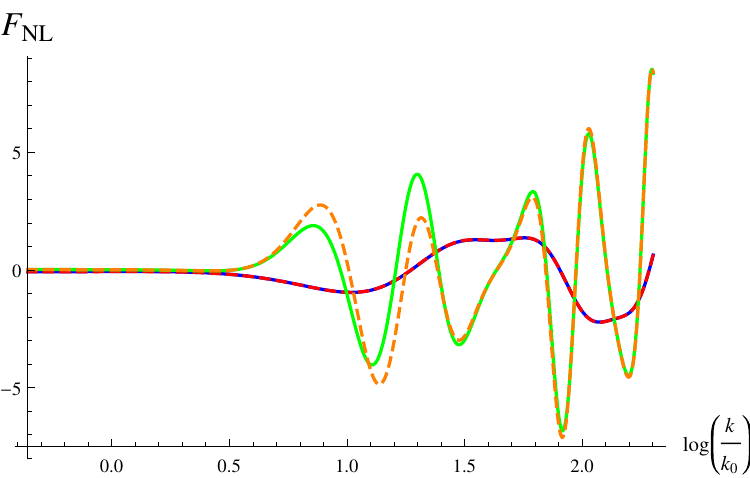}
  \end{minipage}
 \caption{The numerically computed $F_{NL}$ in the squeezed 
(blue-red) and equilateral (green-orange) limits are plotted using the potential in Eq.~\eqn{pottanh} for the parameters in which the spectrum is the same. On the left we use the parameters $\sigma=10^{-3}$ and $n=3$, $\lambda=-8\times 10^{-19}$ (blue and green) and  $n=4$, $\lambda=-6\times 10^{-20}$ (red and orange). On the right we use $\sigma=10^{-4}$ and $n=3$, $\lambda=-2.0\times 10^{-18}$ (blue and green) and  $n=4$, $\lambda=-1.5\times 10^{-19}$ (red and orange).}
\label{FNLtogether}
\end{figure}

\begin{figure}
 \begin{minipage}{.45\textwidth}
  \includegraphics[scale=0.6]{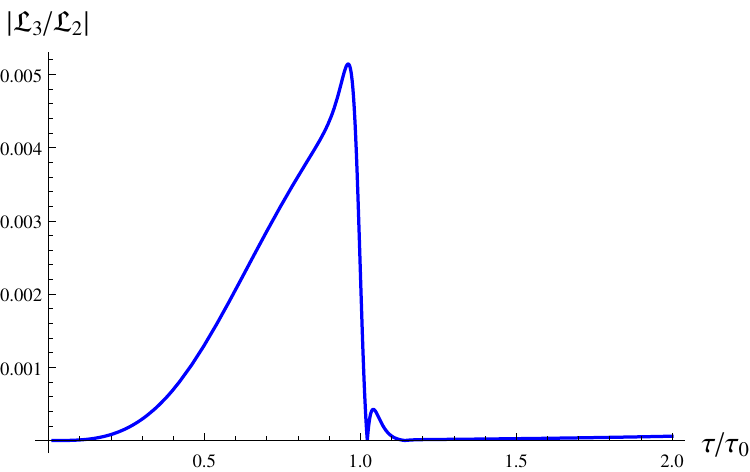}
  \end{minipage}
 \begin{minipage}{.45\textwidth}
  \includegraphics[scale=0.6]{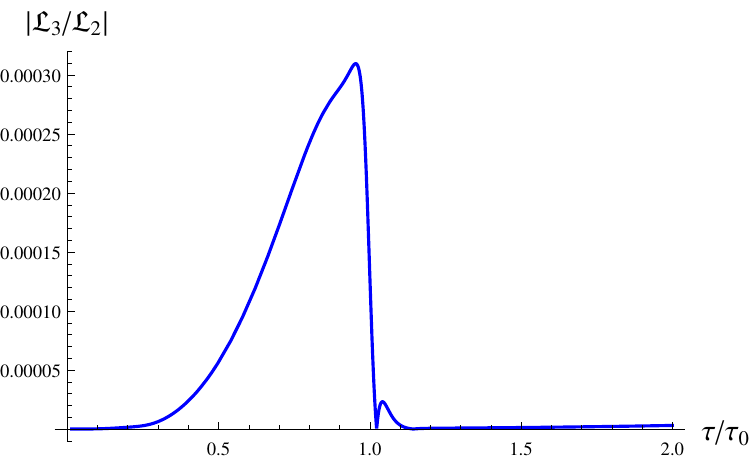}
  \end{minipage}
\begin{minipage}{.45\textwidth}
  \includegraphics[scale=0.6]{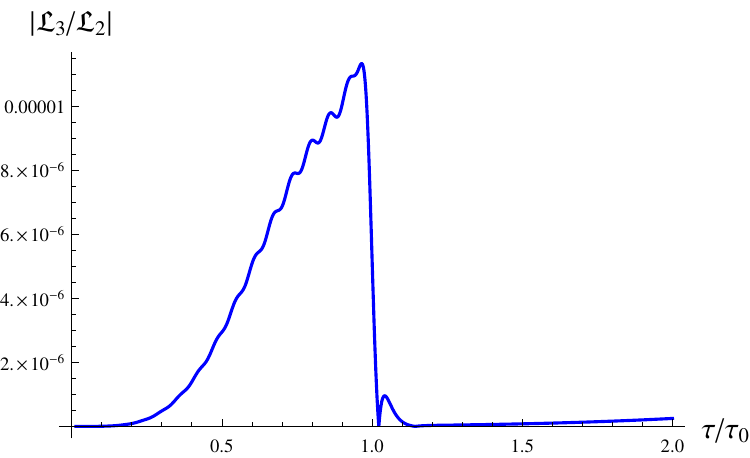}
  \end{minipage}
 \begin{minipage}{.45\textwidth}
  \includegraphics[scale=0.6]{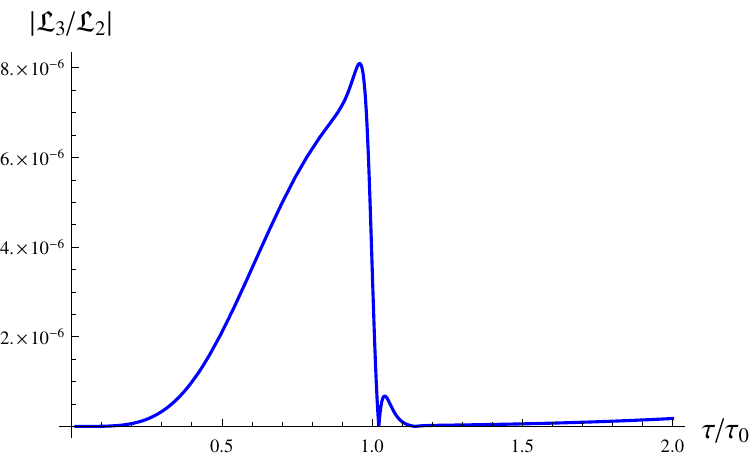}
  \end{minipage}
 \caption{The numerically computed $\mathcal{L}_3/\mathcal{L}_2$ is plotted using the potential in Eq.~\eqn{pottanh} for different scales as a function of conformal time and for the parameters $n=4$ and $\lambda=-1.5\times 10^{-19}$. From left to right and top to bottom we use $k/k_0= 0.1,1, 50$, and $100$.}
\label{l3l2}
\end{figure}

It is important to notice that while  models with sharp features can have some temporary violation of the slow-roll regime, this does not necessarily affect the validity of the effective field theory (EFT) of inflation, since no slow roll approximation is used in deriving the cubic and quadratic action  as was pointed out for example in Refs.~\cite{m,a3}. In order to check the validity of the EFT for the models we have studied  we have plotted in  Fig.~\ref{l3l2} the ratio between the cubic and quadratic actions $|\mathcal{L}_3/\mathcal{L}_2|$  
\begin{eqnarray}
\mathcal{L}_2 &=& a \epsilon ( \mathcal{R}'_c{}^2 - k^2\mathcal{R}_c^2), \\
\mathcal{L}_3 &=& - a \epsilon \eta ( \mathcal{R}_c \mathcal{R}'_c{}^2 + \frac{1}{2} k^2 \mathcal{R}_c^3).
\end{eqnarray}
It can be seen from the plots that for different scales around  $k_0$ the perturbative hierarchy on which EFT is based is not violated, i.e. $|\mathcal{L}_3/\mathcal{L}_2| \ll 1$.

\section{Conclusions}\label{conclusions}
We have studied the degeneracy of the primordial spectrum and bispectrum of the primordial curvature perturbation in single field inflationary models with a class of features of  the inflaton potential. The features  consist in a discontinuous change in the shape of the potential controlled by a couple of parameters that describe the strength of the discontinuity and the change in the potential shape. The feature produces oscillations of the spectrum and bispectrum around the comoving scale $k=k_0$ that exits the horizon when the inflaton passes the discontinuity.
The effects on the spectrum and almost all configurations of the bispectrum including the squeezed limit depend on a single quantity which is a function of the two parameters defining the feature. As a consequence a degeneracy is produced, i.e. different features of the inflaton potential can produce the same observational effects. The degeneracy in the bispectrum is only broken in the equilateral limit around $k_0$. The breaking of the degeneracy in the equilateral limit around $k_0$ could produce an observational signature in the CMB data which could be used to distinguish between different models predicting the same spectrum and bispectrum in other limits and configurations.  We have shown that the degeneracy is also present when considering a continuous potential and that it is only broken in the equilateral limit. This shows that the degeneracy breaking is not an artifact of the non-smoothness of the potential. The breaking of the degeneracy is due to the fact that while $\epsilon$ is approximately the same, $\eta$ can be different.
In the future it will be interesting to obtain an analytic approximation for equilateral limit bispectrum around $k_0$ to better understand  why the degeneracy breaking only occurs in that configuration. Comparison with observations will allow to establish  which modification of the inflaton potential is in better agreement with observational data. It would also be interesting to study these models within the framework of the effective theory of inflation \cite{Cheung:2007st} to understand if this kind of degeneracy could indeed be more general and occur for other inflationary scenarios, not only for single field minimally coupled models.

\acknowledgments
The work of A.G.C. was supported by the Colombian Department of Science, Technology, and Innovation COLCIENCIAS research Grant No. 617-2013. A.G.C. acknowledges the partial support from the International Center for Relativistic Astrophysics Network ICRANet during his stay in Italy. A.G.C. and AER thank the YITP for the kind hospitality during their visit in Kyoto.
%This work was supported by the European Union (European Social Fund, ESF) and Greek national funds under the “ARISTEIA II” Action and the Dedicacion exclusica and Sostenibilidad programs at UDEA, the UDEA CODI projects IN10219CE and 2015-4044.
The work of M.S. was supported by the MEXT KAKENHI Grant Nos.~15H05888 and 15K21733. This work was supported by the UDEA Dedicación exclusiva and Sostenibilidad programs and the CODI projects 2015-4044, 2016-10945, and 2016-13222.

\begin{appendix}
\section{} \label{A1}
The $\mathcal{A}_i$ functions are \cite{Romano:2014kla}
\bea\label{antiderivatives}
&&\mathcal{A}_i(\tau,k_1,k_2,k_3,q)= \frac{(-1)^i(k_2 k_3)^{2(2-i)} H^{3-q}}{\left(4 \epsilon_0 k_1 k_2 k_3\right)^{3/2}} \times \\ \nonumber
\Biggl\{ 
\alpha^*_{k_1} \biggl[ && \alpha^*_{k_2} \Bigl( \mathcal{B}_i(\tau,k_1,k_2,k_3,q) \alpha^*_{k_3} - \mathcal{B}_i(\tau,k_1,k_2,-k_3,q) \beta^*_{k_3} 
\Bigr) \\ \nonumber
+ && \beta^*_{k_2} \Bigl( -\mathcal{B}_i(\tau,k_1,-k_2,k_3,q) \alpha^*_{k_3} + \mathcal{B}_i(\tau,k_1,-k_2,-k_3,q) \beta^*_{k_3} \Bigr) \biggr] \\ 
\nonumber
+ \beta^*_{k_1} \biggl[ && \beta^*_{k_2} \Bigl( \mathcal{B}^*_i(\tau,k_1,k_2,k_3,q) \beta^*_{k_3} - \mathcal{B}^*_i(\tau,k_1,k_2,-k_3,q) 
\alpha^*_{k_3} \Bigr) \\ \nonumber
+ && \alpha^*_{k_2} \Bigl( -\mathcal{B}^*_i(\tau,k_1,-k_2,k_3,q) \beta^*_{k_3} + \mathcal{B}^*_i(\tau,k_1,-k_2,-k_3,q) \alpha^*_{k_3} \Bigr) \biggr]
\Biggr\} ,
\eea
where
\bea
&& \mathcal{B}_1=(\mathrm{i}k_T )^{q-4} \Bigl(k_T \Gamma (3-q,-\mathrm{i} \tau k_T ) +k_1 \Gamma (4-q,-\mathrm{i}\tau k_T ) \Bigr)  \, , \\ \nonumber
&& \mathcal{B}_2= (\mathrm{i}k_T )^{q-4} \Bigl[ k_T^3 \Bigl( \Gamma (1-q,-\mathrm{i} \tau k_T ) + \Gamma (2-q,-\mathrm{i} \tau k_T ) \Bigr) \\ 
&&+ k_T \sum_{i \ne j}^{3}k_ik_j \Gamma (3-q,-\mathrm{i} \tau k_T ) +k_1 \Gamma (4-q,-\mathrm{i} \tau k_T ) \Bigr] \, , \\
&&k_T=k_1+k_2+k_3 \,, \nonumber
\eea
and the $\Gamma$ denotes the incomplete gamma functions defined by
\bea
\Gamma(r,x)= \int_x^{\infty} t^{r-1} e^{-t} dt  \,.
\eea

\end{appendix}

\bibliography{Bibliography}
\bibliographystyle{h-physrev4}
\end{document}